\documentclass[  
a4paper,reprint,   
superscriptaddress,
groupedaddress,  
nofootinbib,
bibnotes,
amsmath,amssymb, 
aps, 
floatfix]{revtex4-1} 

\usepackage{graphicx}
\usepackage{dcolumn}
\usepackage{bm}
\usepackage[mathlines]{lineno}
\usepackage{csquotes}
\usepackage{gensymb}
\usepackage{rotating}
\newcommand\ddfrac[2]{\frac{\displaystyle #1}{\displaystyle #2}}
\usepackage[dvipsnames]{xcolor}
\usepackage{multirow}
\usepackage{siunitx}
\usepackage{comment}

\begin{document}


\title{Composition determination of cosmic rays from the muon content of the showers}
\author{A.~C.~Cobos}
\email{agustin.cobos@iteda.cnea.gov.ar}
\affiliation{Instituto de Tecnolog\'ias en Detecci\'on y Astropart\'iculas Mendoza (CNEA, CONICET, UNSAM),
CNEA Regional Cuyo, Godoy Cruz, Mendoza, Argentina\\}

\author{A.~D.~Supanitsky}
\affiliation{Instituto de Tecnolog\'ias en Detecci\'on y Astropart\'iculas (CNEA, CONICET, UNSAM),
Centro At\'omico Constituyentes, San Mart\'in, Buenos Aires, Argentina\\}

\author{A.~Etchegoyen}
\affiliation{Instituto de Tecnolog\'ias en Detecci\'on y Astropart\'iculas (CNEA, CONICET, UNSAM),
Centro At\'omico Constituyentes, San Mart\'in, Buenos Aires, Argentina\\}

\date{\today}

\begin{abstract}

The origin and nature of ultra high energy cosmic rays remains being a mystery. However, great progress has been made in recent 
years due to the observations performed by the Pierre Auger Observatory and Telescope Array. In particular, it is believed that 
the composition information of the cosmic rays as a function of the energy can play a fundamental role for the understanding of 
their origin. The best indicators for primary mass composition are the muon content of extensive air shower and the atmospheric 
depth of the shower maximum. In this work we consider a maximum likelihood method to perform mass composition analyses based on 
the number of muons measured by underground muon detectors. The analyses are based on numerical simulations of the showers. The 
effects introduced by the detectors and the methods used to reconstruct the experimental data are also taken into account through
a dedicated simulation that uses as input the information of the simulated showers. In order to illustrate the use of the method, 
we  consider AMIGA (Auger Muons and Infill for the Ground Array), the low energy extension of the Pierre Auger Observatory that 
directly measures the muonic content of extensive air showers.  We also study in detail the impact of the use of different high 
energy hadronic interaction models in the composition analyses performed. It is found that differences of a few percent between 
the predicted number of muons have a significant impact on composition determination. 

\end{abstract}

\maketitle


\allowdisplaybreaks 

\section{Introduction}

The cosmic ray energy spectrum extends over more than eleven orders of magnitude in energy (from below $10^{9}$ to above $10^{20}$ eV). It can be approximated by a broken power law with some spectral features: the knee at a few $10^{15}$ eV \cite{Kuli:58,Agli:04,Anto:05,Apel:09,Amen:11}, a second knee at $\sim 10^{17}$ eV \cite{Apel:11}, the ankle at $\sim 5 \times 10^{18}$ eV \cite{Fenu:17}, and a suppression at $\sim 4 \times 10^{19}$ eV \cite{Fenu:17, Iva:17}. Depending on the energy range under consideration, different experimental techniques have been used for the observation of the cosmic rays. Due to their low flux at energies $\gtrsim 10^{15}$ eV, their detection can only be achieved by measuring extensive air showers (EAS), cascades of billions of secondary particles resulting from the interaction of the primary cosmic rays with molecules of the Earth's atmosphere. The EAS present two main components: the electromagnetic one which is formed by electrons, positrons, and  gamma rays, and the muonic one which is formed by muons and antimuons.

Constructed in the province of Mendoza, Argentina, the Pierre Auger Observatory \cite{Aab:15} is the 
largest observatory at present for measuring ultra high energy cosmic rays (UHECR, with energies $\gtrsim 
10^{18}$ eV). This observatory combines arrays of surface detectors (water-Cherenkov tanks) with 
fluorescence telescopes. The first allows one to reconstruct the lateral development of the showers by 
detecting secondary particles that reach the ground. Fluorescence telescopes are used to study the 
longitudinal development of the showers. The combination of the two techniques into a hybrid observatory 
maximizes the precision in the reconstruction of the EAS properties and minimizes systematic errors. 
Located in Utah, USA, the Telescope Array Observatory \cite{Fukushima:03} is also a hybrid detector 
that combines arrays of surface detectors with fluorescence telescopes, in this case the surface 
detectors are composed of scintillator detection devices housed inside metal clad containers. 

Despite great theoretical and experimental efforts done in recent years, the cosmic ray origin still 
remains a mystery.  Recent results \cite{Risse:07,Abra:07,Abra:10} suggest that the UHECR flux is 
composed predominantly of hadronic primary particles. As charged particles, they suffer deflections in 
cosmic magnetic fields and their directions do not point back directly to their sources. Therefore, an 
indirect search for their origin is necessary: the measurement of the energy spectrum, the estimation of 
primary mass composition as a function of the energy, and the distribution of their arrival directions. 
In particular, composition information appears to be crucial to find the transition between the galactic 
and extragalactic components of the cosmic rays \cite{Medi:07} and to elucidate the origin of the 
suppression at the highest energies \cite{Kamp:13}.  
 
Together with the atmospheric depth corresponding to the maximum shower development, $X_{max}$, the best indicator of primary mass composition is the muon content of the shower \cite{Kamp:12, Supa:08}. In fact, heavier primaries produce more muons than lighter ones. The Auger Muon and Infill for the Ground Array (AMIGA) is an extension of the Pierre Auger Observatory that directly measures the muonic content of EAS \cite{Ravi:16}. It will consist in two triangular grids of $750$ m and $433$ m spacing composed by pairs of detectors, a water-Cherenkov tank and a $30$ $\textrm{m}^2$ muon counter buried underground. AMIGA operates in the energy region from $\sim 10^{16.5}$ to $\sim 10^{19}$ eV. With sufficient statistics, AMIGA will contribute to the mass composition determination in this energy range. 

Primary mass composition analyses can only be performed by comparing experimental data with EAS simulations. These simulations are subject to large systematic uncertainties because they are based on high energy hadronic interaction models (HEHIMs) that extrapolate low energy accelerator data to the highest energies. The most used HEHIMs in the literature have been recently updated  by using data taken by the Large Hadron Collider (LHC). These models (Sibyll 2.3c \cite{Riehn:17}, EPOS-LHC \cite{Pierog:15}, and QGSJETII-04 \cite{Ostap:13}) are called post-LHC models, due to their tuning to LHC data. Concerning the number of muons at ground, predictions of these HEHIMs differ only by about 10\% \cite{Pierog:17}. However, experimental results indicate that the muon content of the showers is 30 to $80 \%$ greater than that estimated from simulations \cite{Aab2:15, Aab:16, Muller:18, Dembinski:19}. A parameter very closely related to the muon content of the showers is the muon density at a given distance to the shower axis, which presents a dependence on the zenith angle of the EAS \cite{Supa:09, Apel:17}.  

In this work, a Maximum Likelihood method developed to perform primary mass composition analyses is considered. Here, the parameter 
sensitive to primary mass is the number of muons detected at ground at a given distance to the shower axis and for different zenith 
angle of EAS. The studies are performed by using numerical simulations, which include experimental uncertainties in the reconstruction 
of the energy and in the measurement of the number of muons. The effect of the shape of the cosmic ray energy spectrum is also considered. 
The analyses are performed for binary mixtures of different hypothetical values of proton primary abundance. The method combines 
all values of the number of muons in a given zenith angle range. The impact of the differences between HEHIMs predictions of the number 
of muons at ground as a function of the zenith angle is also studied. It is worth mentioning that in this work several parameters of 
the AMIGA design are assumed but the same study can be applied to any other experiment that involves muon number measurements.

\section{Analysis}\label{secAnalysis}

\subsection{Maximum Likelihood Method}\label{sec1}
\label{ssec:MaxLikeMet}

In this section the Maximum Likelihood (ML) method to  determine the mass composition is described. As mentioned in the previous section, the composition analysis is carried out based on the number of muons at ground for the same distance to the shower axis. Therefore, all shower variables and distribution functions defined hereafter will be referred to this fixed parameter.

Let $\rho_{\mu}(\theta)$ be the muon density of a shower with zenith angle, $\theta$. The number of
muons, $N_{\mu}$, that impact to a horizontal muon counter of area $a_d$, is computed as 
\begin{equation}
\label{eq:PN}
N_{\mu}  =  \rho_{\mu}(\theta) \ a_d \cos(\theta).
\end{equation}

Let $P^{\theta}_A(\widetilde{N}_{\mu}|E) \equiv P(\widetilde{N}_{\mu}| E,\textrm{sec}(\theta), A)$ be 
the distribution function of the measured (reconstructed) number of muons, $\widetilde{N}_{\mu}$, due 
to a primary of type $A$ with  a zenith angle $\theta$ and energy $E$. Whereas the number of muons is 
a function of the true energy $E$, the measured number of muons is a function of the reconstructed 
energy $E_R$. Then, the probability 
$P^{\theta}_A(\widetilde{N}_{\mu}|E_{Ri}) \equiv P(\widetilde{N}_{\mu}| E_{Ri},\textrm{sec}(\theta), A)$ 
of $\widetilde{N}_\mu$ calculated in the $i$-th reconstructed energy bin, takes the following form (see 
appendix A of Ref.~\cite{Supa:08}),
\begin{widetext}
\begin{equation}
P^{\theta}_A(\widetilde{N}_{\mu}|E_{Ri}) = \ddfrac{\int_{0}^{\infty} \int_{E_{Ri}^{-}}^{E_{Ri}^{+}}
 J(E)\, G(E_{R}|E)\,  P^{\theta}_A(\widetilde{N}_{\mu}|E)\, dE\, dE_{R}}{ \int_{0}^{\infty} \int_{E_{Ri}^{-}}^{E_{Ri}^{+}} J(E)\, G(E_{R}|E)dE\, dE_{R}},
\label{EqRhoAv}
\end{equation}
\end{widetext}
where $E_{Ri}$ is the center of the $i$-th reconstructed energy bin, $E_{Ri}^{-}$ and $E_{Ri}^{+}$ are
the lower and upper limits of that bin, $J(E)$ is the cosmic ray energy spectrum, and $G (E_{R}|E)$ is
the conditional probability distribution of $E_R$ conditioned to $E$.  

Note from Eq.~(\ref{EqRhoAv}) that the energy of a real or simulated air shower with true energy $E$ 
is estimated by means of the reconstruction procedure producing a value, $E_R$, according to $G(E_R|E)$. 
Furthermore, the distribution of the true energy $E$ is given by the cosmic ray spectrum $J(E)$. 

For the composition method described in this section let us consider the simplified case in which there are just two nuclear species, $A_1$ and $A_2$. Let $N$ be the number of detected showers, i.e.~$N$ is the sample size. Then, the probability of the configuration 
$\mathbf{\widetilde{N}} = (\widetilde{N}_{\mu,1}, ..., \widetilde{N}_{\mu,N})$ is given by,
\begin{eqnarray}
P(\mathbf{\widetilde{N}}\, |E_{Ri}, c_{A_1}) &=&  \prod^{N}_{j=1} \left[c_{A_1} P^{\theta_j}_{A_1}(\widetilde{N}_{\mu,j}|E_{Ri})  \right. \nonumber \\
&& \left. + \; (1-c_{A_1})\, P^{\theta_j}_{A_2}(\widetilde{N}_{\mu,j}|E_{Ri}) \right], 
\label{eq:PN}
\end{eqnarray}
where $c_{A_1}$ is the abundance of $A_1$. Taking the logarithm of Eq.~(\ref{eq:PN}) and equating to zero its derivative with respect to $c_{A_1}$, the following condition for the estimator of $c_{A_1}$, $\hat{c}_{A_1}$, is obtained
\begin{equation}
\label{eq:ML}
\sum^{N}_{j=1}\frac{P^{\theta_j}_{A_1}(\widetilde{N}_{\mu,j}|E_{Ri})-P^{\theta_j}_{A_2}(\widetilde{N}_{\mu,j}|E_{Ri})}{\hat{c}_{A_1} P^{\theta_j}_{A_1}(\widetilde{N}_{\mu,j}|E_{Ri}) + (1-\hat{c}_{A_1})P^{\theta_j}_{A_2}(\widetilde{N}_{\mu,j}|E_{Ri})}=0.
\end{equation}
Therefore, the solution of Eq.~(\ref{eq:ML}) gives the maximum likelihood estimator of the abundance
of the $A_1$ nuclear type.

\subsection{Simulations of EAS and $P^{\theta}_A(\widetilde{N}_{\mu}|E_{Ri})$ determination}
\label{subsec:EASsim}

In order to calculate the distribution functions considered in this work, different EAS simulations were performed. The shower library used in this work is generated with CORSIKA v76300 \cite{Knapp:98}. The HEHIMs considered are EPOS-LHC \cite{Pierog:15} and Sibyll 2.3c \cite{Riehn:17}. Proton ($A_1 = p$) and iron ($A_2 = Iron$) are considered as primaries in sections \ref{sec:perform} and \ref{sec:currvsfut}, while nitrogen ($A_2 = Nitrogen$) is considered in section \ref{sec:simpcase}. The low-energy hadronic interactions are  simulated by using FLUKA \cite{Ferr:05}. The ground level is set at the Auger altitude ($1452$ m). The magnetic field at the Auger location is taken into account. The showers are simulated for primary energies between $10^{17.25}$ and $10^{18.75}$ eV in steps of $\Delta \log(E/\textrm{eV})=0.25$ and zenith angle, $\theta$, corresponding to $\sec(\theta)$ between 1 and $1.5$ in steps of $0.1$. For each  direction, a set of 100, 30 and 10 EAS are generated for proton, nitrogen, and iron, respectively. Muons at 450 m from the shower axis (sampled in a 20 m wide ring) are considered since this is the distance that minimizes the fluctuations for a 750 m array spacing \cite{Newton:07, Ravignani:15}. Muon counters with $a_d = 30\ \textrm{m}^2$, $100 \%$ of efficiency, and buried underground at $2.3$ m depth, which corresponds to a muon energy threshold of 1 GeV for a vertical incidence, are considered. That is, only muons with energy greater than 1 GeV/cos($\theta_{k}$) reach the detector, being $\theta_{k}$ the zenith angle of the direction of motion of the individual muons. 

\begin{figure}[t]
\centering
\setlength{\abovecaptionskip}{0pt}
\includegraphics[width=\linewidth]{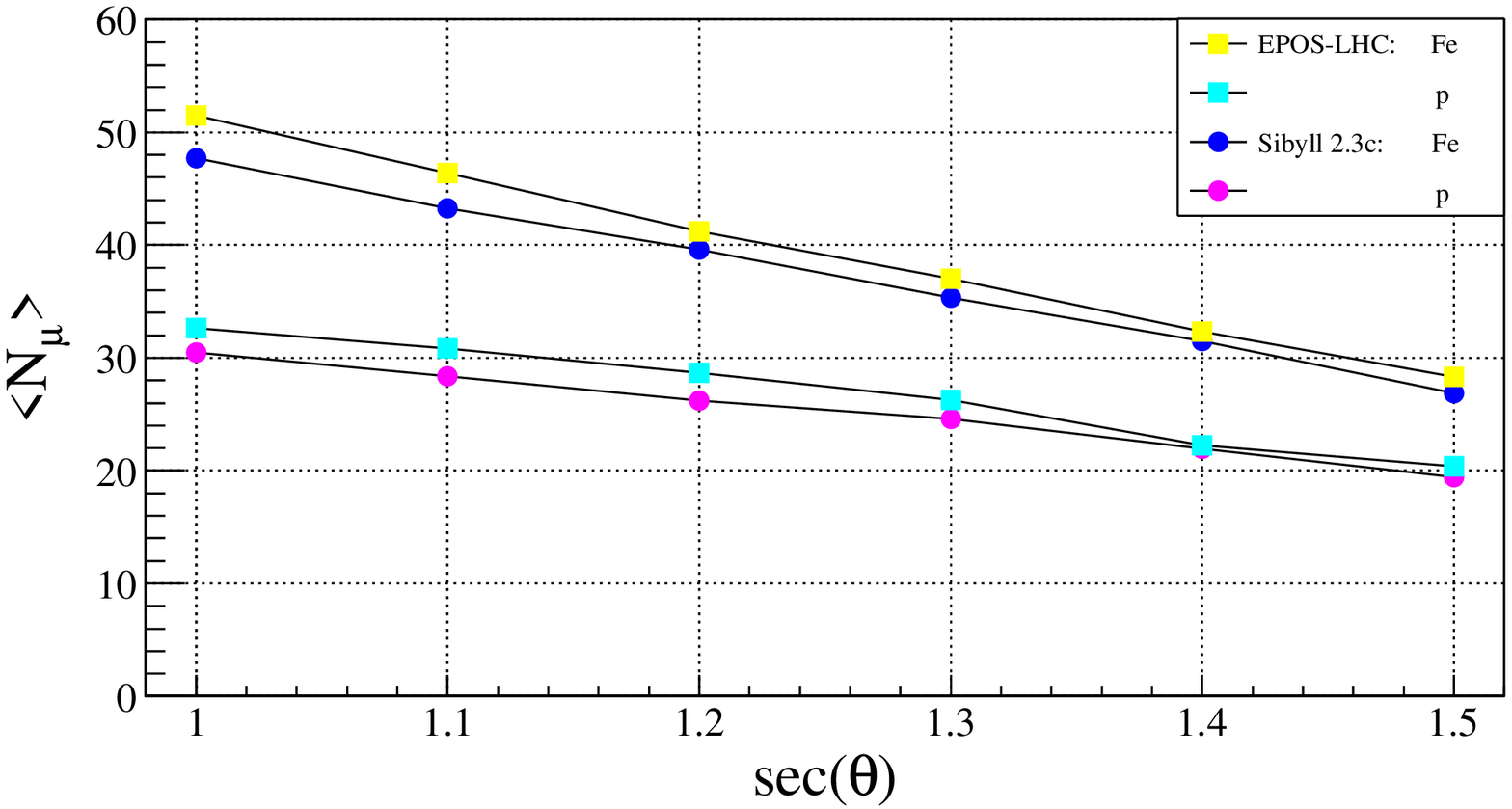}
\includegraphics[width=\linewidth]{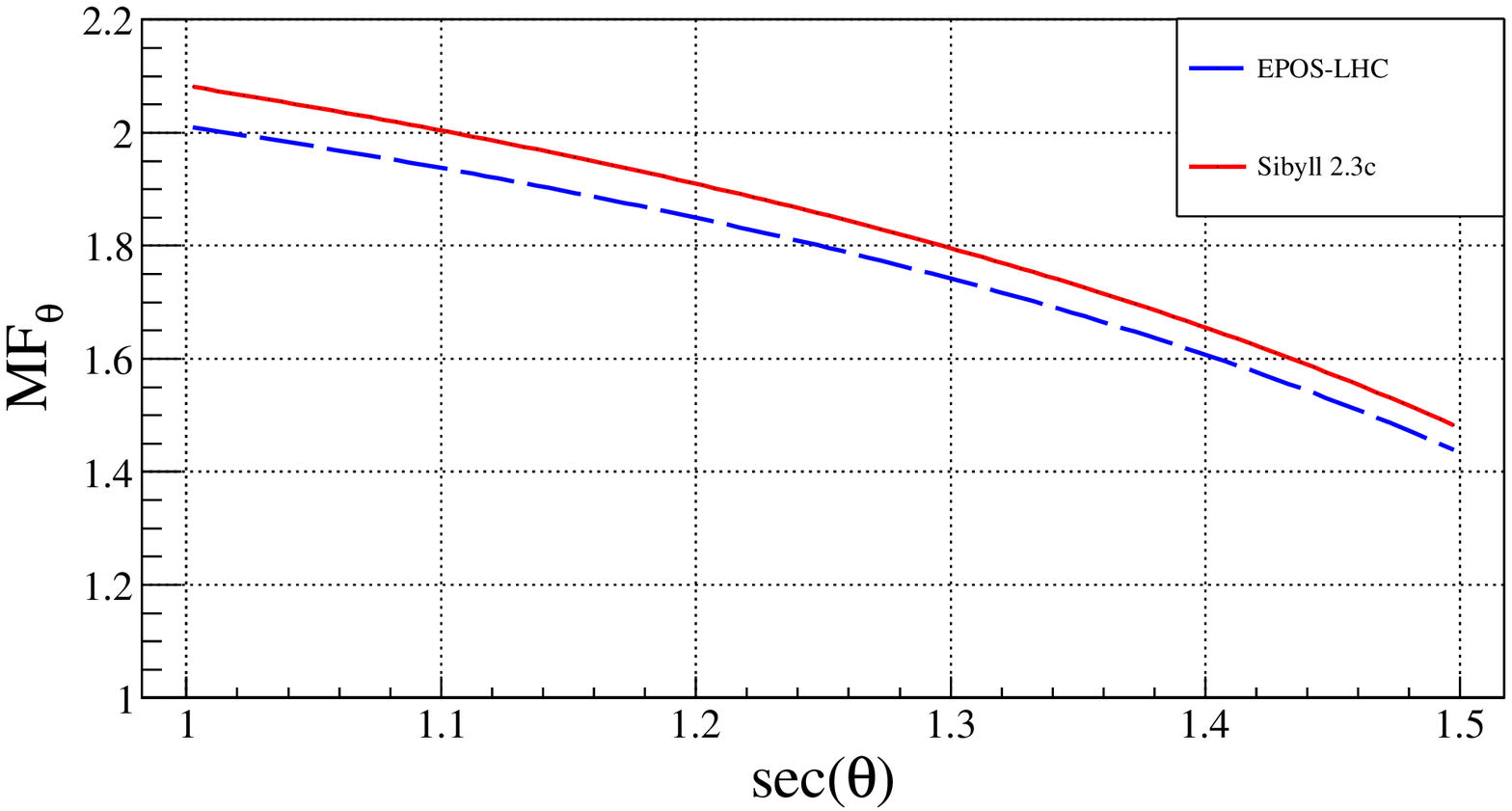}
\caption{Top: $\langle N_\mu \rangle$ as a function of sec$(\theta)$ for $E=10^{18}$ eV corresponding 
to proton ($p$) and iron ($Fe$) primaries. The error bars are smaller than the marker size. Bottom: 
Merit factor of $\widetilde{N}_{\mu}$ as a function of $\sec(\theta)$. The two HEHIMs considered are 
EPOS-LHC and Sibyll 2.3c.\label{MF}}
\end{figure}

The number of muons at a given distance from the shower axis presents shower to shower fluctuations.
Its distribution function is characterized by the mean value 
$\langle N_\mu \rangle^{\theta,E}_A \equiv \langle N_\mu \rangle(E,\textrm{sec}(\theta), A)$ and the 
standard deviation, 
$\sigma_{sh}[N_\mu]^{\theta,E}_{A} \equiv \sigma_{sh}[N_\mu](E,\textrm{sec}(\theta), A)$. It is 
worth mentioning that the distribution functions present asymmetric tails, which is commonly
found in EAS physics. The top panel of Fig.~\ref{MF} shows the mean value of the number of muons 
as a function of $\sec(\theta)$ for proton and iron primaries with $E=10^{18}$ eV and for the two 
HEHIMs considered. It can be seen that the mean value of $N_\mu$ is a decreasing function of 
$\sec(\theta)$ and also that, as mentioned before, the differences between the predictions 
corresponding to the two HEHIMs considered are of the order of $10\%$ for proton and iron primaries.   

As mentioned above, the distribution function of $N_\mu$ presents asymmetric tails. However, the
distribution function of the reconstructed $N_\mu$, i.e.~$\widetilde{N}_{\mu}$, is given by the 
convolution of the distribution function of $N_\mu$ with the one that takes into account the 
fluctuations introduced by the detectors and the effects of the reconstruction methods. As a result,
a Gaussian distribution is a good approximation of the distribution function corresponding to the 
reconstructed number of muons \cite{Supa:15}, which is given by,
\begin{equation}
\label{P_A}
 P^{\theta}_A(\widetilde{N}_{\mu}|E) = \frac{1}{\sqrt{2 \pi}\ \sigma[\widetilde{N}_\mu]^{\theta,E}_A} \exp\left[ -\frac{\left(\widetilde{N}_\mu-\langle N_\mu \rangle^{\theta,E}_A\right)^2}{2\ \sigma^2[\widetilde{N}_\mu]^{\theta,E}_A}\right], 
\end{equation}
where 
$\sigma[\widetilde{N}_\mu]^{\theta,E}_A \equiv \sigma[\widetilde{N}_\mu](E,\textrm{sec}(\theta), A)$ is
given by,
\begin{eqnarray}
\sigma^2[\widetilde{N}_\mu]^{\theta,E}_A &=& \sigma_{sh}^{2}[N_\mu]^{\theta,E}_A \left(
\sigma^{2}[\epsilon]^{\theta,E}_A + 1 \right) \nonumber \\
&& + \sigma^{2}[\epsilon]^{\theta,E}_A \left[\langle N_\mu \rangle^{\theta,E}_A\right]^2.
\label{sig_A}
\end{eqnarray}
Here $\sigma[\epsilon]^{\theta,E}_A \equiv \sigma[\epsilon](E,\sec(\theta),A)$ is the relative error 
of the reconstructed number of muons, 
i.e.~$\sigma[\epsilon]=\sigma[\widetilde{N}_\mu/\langle N_\mu \rangle -1]$. 

\begin{figure}[ht!]
\centering
\setlength{\abovecaptionskip}{0pt}
\includegraphics[width=\linewidth]{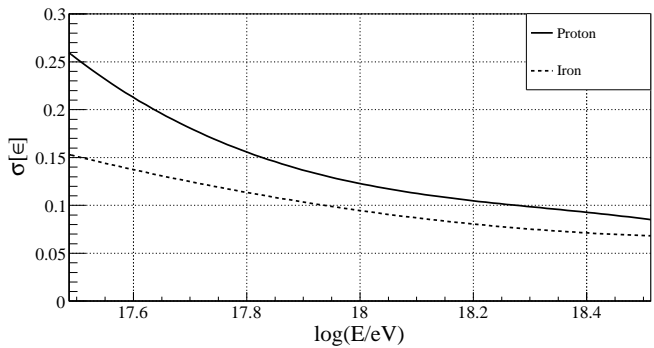}
\caption{Relative error corresponding to the reconstructed number of muons as a function of the logarithm 
of the energy for proton and iron nuclei and zenith angle $\theta=45^\circ$.
\label{rec}}
\end{figure}

In Ref.~\cite{Supa:15} $\sigma[\epsilon]$ at 750 m from the shower axis is calculated from 
simulations of the showers and the AMIGA detectors by using the reconstruction method of the muon 
lateral distribution function developed in that work. In that calculation the core position and the
arrival direction of the showers are reconstructed by using the information of the water Cherenkov 
detectors whereas the muon lateral distribution function is reconstructed by using the information 
given by the muon counters but using the core position determined with the Cherenkov detectors. 
The values of $\sigma[\epsilon]$ used in this work are obtained by fitting the data points shown in 
Fig.~9 (right panel) of Ref.~\cite{Supa:15} between $10^{17.6}$ and $10^{18.5}$ eV, for proton and 
iron primaries and for $\theta=45^\circ$. The proton data points are fitted with a cubic function of 
$\log(E/\textrm{eV})$ with coefficients: $a_0 = 1163.0$,  $a_1 = -190.199$, $a_2 = 10.3725$ and 
$a_3 = -0.188612$. The iron data points are fitted with a quadratic function of $\log(E/\textrm{eV})$ 
with coefficients: $b_0 = 21.2686$,  $b_1 = -2.27008$ and $b_2 = 0.0607638$. Note that $a_i$ and $b_i$ 
are the coefficients of the $i$th power of $\log(E/\textrm{eV})$. Fig.~\ref{rec} shows the fitted
$\sigma[\epsilon]$ as a function of $\log(E/\textrm{eV})$ for proton and iron primaries. Although 
$\sigma[\epsilon]$ is nearly independent of zenith angle, the $\theta=45^\circ$ data points are
considered due to their slightly larger values compared with the ones obtained for $\theta=30^\circ$ 
(see Fig.~9 of Ref.~\cite{Supa:15}). Also, $\sigma[\epsilon]$ at 750 m from the shower axis is used 
for $\sigma[\epsilon]$ at 450 m, this is an approximation based on the results obtained in 
Ref.~\cite{Ravignani:15} in which it is shown that a similar or even smaller value of $\sigma[\epsilon]$ 
is obtained at 450 m from the shower axis considering an improved reconstruction method developed in 
that work. Therefore, the values of $\sigma[\epsilon]$ used for the present calculations include the 
effects introduced by the detectors and the reconstruction methods conservatively.  

$\langle N_\mu \rangle^{\theta,E}_A$ and $\sigma_{sh}[N_\mu]^{\theta,E}_A$ are obtained from the 
CORSIKA simulations. For each primary type and HEHIM both quantities are obtained by fitting
the simulated data with linear functions of $\sec(\theta)$ such that the coefficients are cubic 
functions of $\log(E/\textrm{eV})$.

The discrimination power of a given mass sensitive parameter, $q$, can be assessed by the commonly used 
merit factor, which is defined as,
\begin{equation}
\textrm{MF}(q)=\frac{\langle q \rangle_{A_2}-\langle q \rangle_{A_1}}{\sqrt{\textrm{Var}[q]_{A_2}+\textrm{Var}[q]_{A_1}}},
\end{equation} 
where $\textrm{Var}[q]_A$ is the variance of parameter $q$ for the primary type $A$. The bottom panel
of Fig.~\ref{MF} shows the $\textrm{MF}(\widetilde{N}_{\mu})$ as a function of $\sec(\theta)$ (denoted 
in this work as $\textrm{MF}_{\theta}$) for the two HEHIMs. It can be seen that the curves present 
similar $\textrm{MF}_{\theta}$ due to the small differences between the number of muons predicted by 
the two different HEHIMs considered. It can also be seen that the curves have $\textrm{MF}_{\theta}$
values compatible with those found in Ref. \cite{Supa:08} reaffirming that the muon content of the 
showers is the best indicator of primary mass composition together with $X_{max}$ 
\cite{Kamp:12, Supa:08}. 

The energy range considered in this work ranges from $10^{17.5}$ to $10^{18.5}$ eV, which corresponds
to the 750 m array of AMIGA. In this energy range the cosmic ray flux can be approximated as
$J(E) =C\, E^{-3.27}$ \cite{Verzi:19} where $C$ is a normalization constant. The conditional
probability distribution $G(E_{R}|E)$ is assumed to be a Gaussian distribution,
\begin{equation}
\label{eq:sig_E}
G (E_{R}|E) = \frac{1}{\sqrt{2 \pi}\ \sigma[E]} \exp\left[ -\frac{(E_R-E)^2}{2\ \sigma^2[E]} \right], 
\end{equation}
where $\sigma[E] = (0.084 + 0.047 \sqrt{10^{18}\, \textrm{eV}/E})\times E$ is the surface detectors energy resolution corresponding to Auger in the energy range under consideration
\cite{Cole:19}. Then, by using Eq.~(\ref{eq:sig_E}) it is possible to express Eq.~(\ref{EqRhoAv}) 
as follows, 
%
%
\begin{equation} 
P^{\theta}_A(\widetilde{N}_{\mu}|E_{Ri}) = \ddfrac{\int_{0}^{\infty} \!\!\!
I(E,E_{Ri}^{+},E_{Ri}^{-})\,J(E)\, P^{\theta}_A(\widetilde{N}_{\mu}|E)\, dE}{ \int_{0}^{\infty}\!\!\!%
I(E,E_{Ri}^{+},E_{Ri}^{-})\, J(E)\, dE},
\label{eq:I}
\end{equation} 
%
%
where 
\begin{equation}
I(E,E_{Ri}^{+},E_{Ri}^{-}) = \frac{1}{2} \left[\text{erf} \left(\frac{E_{Ri}^{+} - E}{\sqrt{2}\, %
\sigma[E]}\right) - \text{erf}\left(\frac{E_{Ri}^{-} - E}{\sqrt{2}\, \sigma[E]}\right) \right], 
\end{equation}
being $\text{erf}(x)$ the error function. Reconstructed energy bins of width 
$\log(E_{Ri}^{+}/\textrm{eV}) - \log(E_{Ri}^{-}/\textrm{eV}) = 0.1$ centered at $E_{Ri}$ are considered. Finally, it is worth mentioning that the integrals of Eq.~(\ref{eq:I}) are solved numerically.

\subsection{Simulations for the study of the proton abundance estimator}
\label{SecSim}

The simulations for the study of the performance of $\hat{c}_p$, the estimator of the proton abundance  ($A_1 = p$), $c_p$, are performed by using the ROOT package \cite{Root}. The values of $c_p$ considered range between
$0.1$ and $0.9$ in steps of $0.1$. Given the $i$-th reconstructed energy bin, for each value of $c_p$, 
the number of events, $N_p$, due to proton induced air showers is obtained by sampling a Binomial 
distribution function, $B(N, c_p)$, where $N$ is the total number of events in the $i$-th bin. $N$
is obtained by sampling a Poisson distribution with mean value given by, 
\begin{equation}
\mu_i = T\, A_c\, \int_{E_{Ri}^-}^{E_{Ri}^+} J(E) dE  
\end{equation}
where $T$ is the observation time considered (5 and 10 years) and $A_c$ is the acceptance of the 
detector. The zenith angle interval from 0 to $45^{\circ}$ is considered (the one corresponding to the 
AMIGA muon detectors), then, $A_c = S\, \pi \sin^2(45^{\circ})$ where $S=25$ $\textrm{km}^{2}$ is 
approximately the area of the 750 m AMIGA array. The number of events due to the other primary $A_2$ 
($Iron$ or $Nitrogen$) is calculated as $N_{A_2} = N - N_p$. 

The simulated energy $E_j$ of an event that belongs to the $i$-th bin is obtained by sampling the 
flux $J(E)$ (adequately normalized) in a wide energy interval centered at $E_{Ri}$. The reconstructed
energy $E_{R,j}$ is obtained by sampling the Gaussian distribution of Eq.~(\ref{eq:sig_E}) centered at
$E_j$. This sampling process is repeated until the reconstructed energy $E_{R,j}$ falls in the $i$-th
reconstructed energy bin. The zenith angle corresponding to each event is taken at random from an
isotropic distribution ($f(\theta) \propto \sin(\theta) \cos(\theta)$) in the zenith angle range 
mentioned before. The parameter $\widetilde{N}_\mu$ for each event is obtained by sampling a Gaussian
distribution (see Eq.~(\ref{P_A})) obtained evaluating the mean value and the standard deviation,
corresponding to the primary type of the event, in the zenith angle and simulated energy obtained 
before.  

For each value of $c_p$, $10^3$ independent samples are generated in order to obtain the distribution 
function of $\hat{c}_p$. The mean value of the proton abundance estimator, $\langle \hat{c}_p \rangle$,
and its standard deviation, $\sigma_{\hat{c}_p}$ are calculated.

In order to analyze the impact of the differences between HEHIMs on composition analyses a given HEHIM, 
called the reference model, is used to generate the event samples, which is not necessarily the same 
as the one used to analyze the data, i.e.~to calculate the parameters of the Gaussian distributions
involved in the calculation of $\hat{c}_p$ by means of Eq.~(\ref{eq:ML}).  

We consider the HEHIMs Sibyll 2.3c, EPOS-LHC, and *Sibyll 2.3c, a modified version of Sibyll 2.3c for
which the values of $\langle N_\mu \rangle^{\theta,E}_{A}$ and $\sigma[\widetilde{N}_\mu]^{\theta,E}_A $ 
are obtained by multiplying the ones corresponding to Sibyll 2.3c by a factor (1 + $\varepsilon$) with $
\varepsilon = \textrm{constant}$. Note that *Sibyll 2.3c is constructed in such a way that the parameter 
$N_\mu$ has the same merit factor as the one obtained for Sibyll 2.3c, regardless the pair of primary 
types considered.

In order to analyze the results obtained from the simulations the following quantities are considered: 
The bias of $\hat{c}_{p}$, which is given by,
\begin{equation}
\delta_{c_{p}} = \langle \hat{c}_{p} \rangle(\textrm{HM},\textrm{HM}_{ref}) - c_{p},
\end{equation}
and the percentage difference, $\Delta \sigma_{c_{p}}$, which is defined as,
\begin{equation}
\Delta \sigma_{c_{p}} =100\%\ \left[ \frac{\sigma_{\hat{c}_{p}}(\textrm{HM},\textrm{HM}_{ref})}{\sigma_{\hat{c}_{p}}(\textrm{HM}_{ref},\textrm{HM}_{ref})}-1 \right],
\end{equation}
where $\langle \hat{c}_{p} \rangle(\textrm{HM},\textrm{HM}_{ref})$ and $\sigma_{\hat{c}_{p}}(\textrm{HM},\textrm{HM}_{ref})$ are the mean value and the standard deviation of the estimator $\hat{c}_{p}$, respectively, $\textrm{HM}_{ref}$ is the reference model considered, and HM is the HEHIM used to analyze the simulated data. 

\section{Results and Discussion} 
\label{secResults}
    
\subsection{Performance of the Maximum Likelihood method}
\label{sec:perform}

\begin{figure}[t!]
\centering
\setlength{\abovecaptionskip}{0pt}
\includegraphics[width=\linewidth]{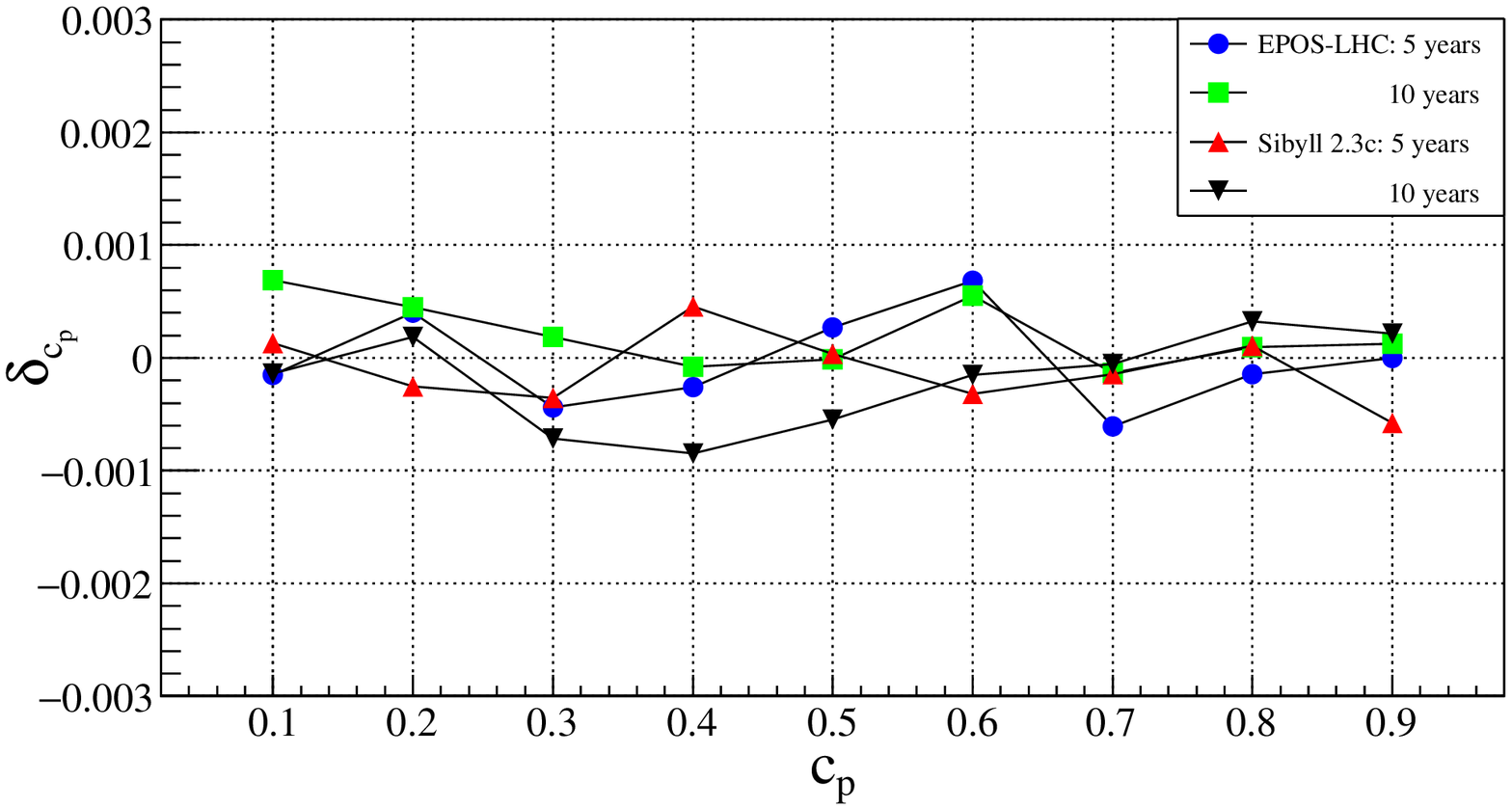}
\includegraphics[width=\linewidth]{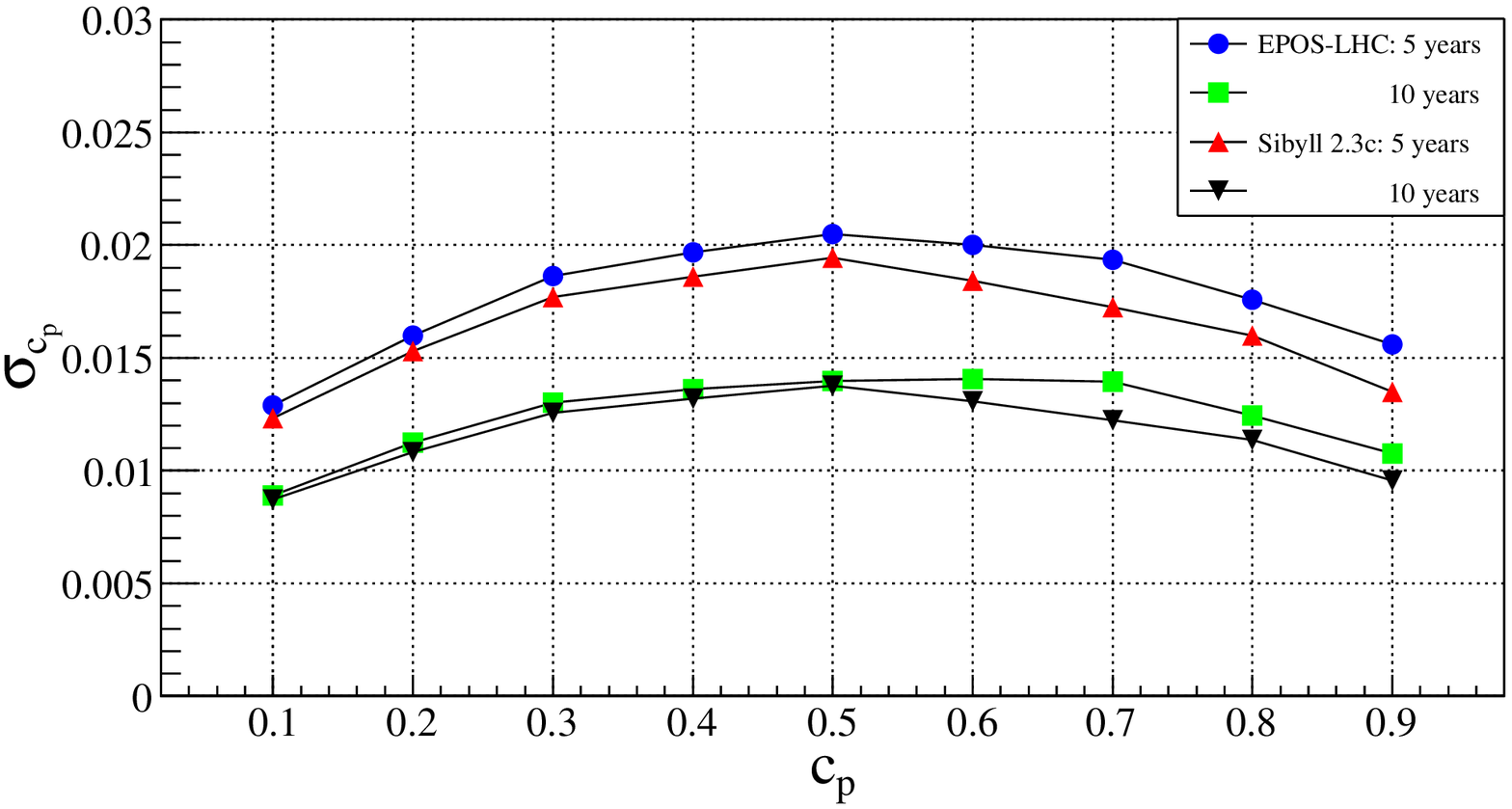}
\caption{$\delta_{c_p}$ (top) and $\sigma_{\hat{c}_p}$ (bottom) versus $c_p$ calculated with the ML method for 
$E_{Ri}=10^{18}$ eV, sample size $N$ corresponding to 5 and 10 years of observation, and for binary mixtures of 
proton and iron primaries. The HEHIMs used are Sibyll 2.3c and EPOS-LHC. In both cases the reference model is 
also used to analyze the data, i.e.~ HM=HM$_{ref}$. \label{sigML}}
\end{figure}

Figure \ref{sigML} shows $\delta_{c_p}$ (top) and $\sigma_{\hat{c}_p}$ (bottom) as a function of $c_p$ 
calculated with the ML method for Sibyll 2.3c and EPOS-LHC used each one of those as the reference model and to analyze 
the data (HM = HM$_{ref}$), $E_{Ri}=10^{18}$ eV and random sample size $N$ corresponding to 5 and 10 years of collected 
events at AMIGA 750 m array ($N \sim 10^{3}$ for 5 years). From Fig.~\ref{sigML} (top) one can see that the bias is 
negligible over the entire $c_p$ range for both HEHIMs considered. It is worth mentioning that in previous studies done 
for fixed values of the zenith angle the biases obtained are also negligible in the entire zenith angle range considered.
From the $\sigma_{\hat{c}_p}$ values of Fig.~\ref{sigML} (bottom) it can also be seen that proton abundance is estimated 
with high resolution showing that the muon content of the shower is one of the best indicators of primary mass composition. 
Fig.~\ref{sigML} (bottom) also shows that the differences between the values of $\sigma_{\hat{c}_p}$ obtained with Sibyll 2.3c 
and EPOS-LHC are less than $10 \%$ due to the similarity of their $\textrm{MF}_{\theta}$ (Fig.~\ref{MF} bottom).
From the figure it can also be seen that $\sigma_{\hat{c}_p}$ has a maximum around $c_p=0.5$ and that it is smaller for 
$c_p=0.1$ than for $c_p=0.9$. The maximum at $c_p \cong 0.5$ can be understood from the fact that at $c_p=0.5$ the 
fluctuations on the number of proton events only and in the number of iron events only are the largest (from the expression
of the standard deviation of a binomial variable it can be seen that $\sigma[N_{p}]=\sigma[N_{Fe}]=\sqrt{N}/2$) causing 
the appearance of the maximum. For values of $c_p$ close to zero the number of iron events increases and its fluctuations 
decrease ($\sigma[N_{Fe}]=\sqrt{N c_p (1-c_p)}$) causing the decrease of the uncertainty on the determination of $c_p$. 
The same happens for values of $c_p$ close to one. The larger values of $\sigma_{\hat{c}_p}$ obtained at $c_p=0.9$ compared 
with the ones obtained for $c_p=0.1$ has to do with the fact that the width of the $\widetilde{N}_\mu$ proton distribution 
is larger than the one corresponding to iron.   

\begin{figure}[t]
\centering
\setlength{\abovecaptionskip}{0pt}
\includegraphics[width=\linewidth]{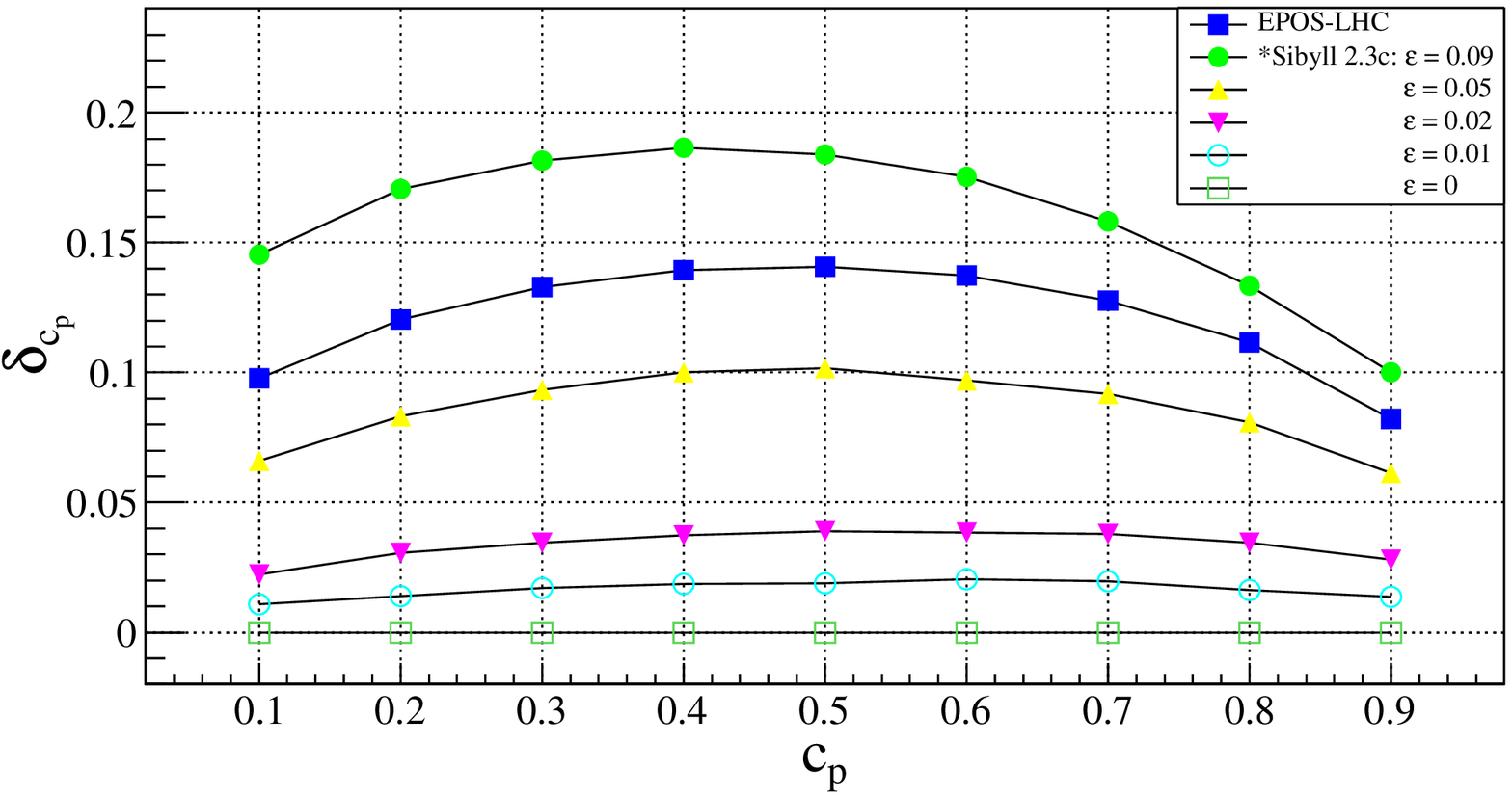}
\includegraphics[width=\linewidth]{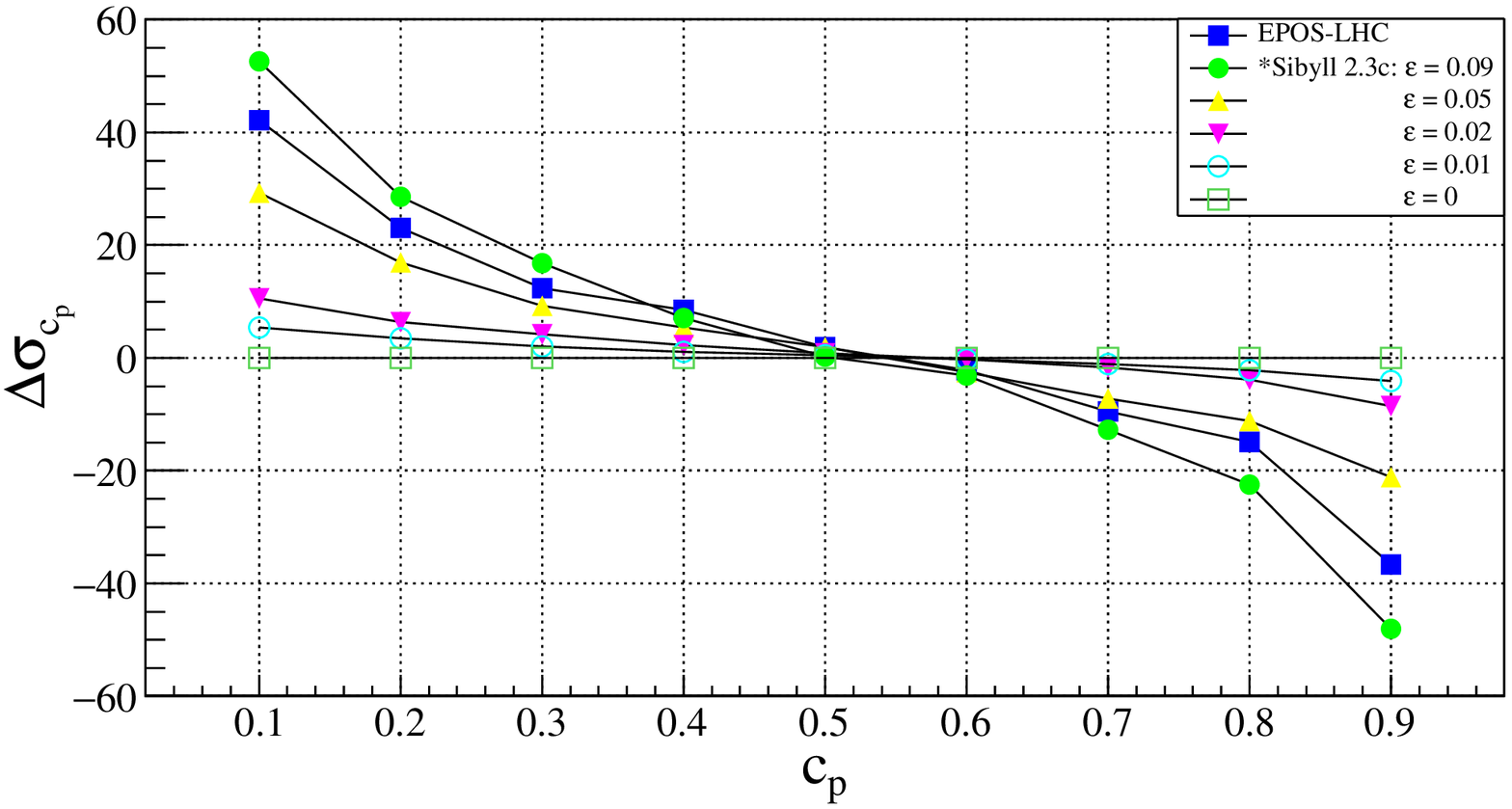}
\caption{$\delta_{c_p}$ (top) and $\Delta \sigma_{c_p}$ (bottom) versus $c_p$ for the ML method and for
different values of $\varepsilon$ (see section \ref{SecSim}). The case corresponding to HM=EPOS-LHC is
also shown. All cases correspond to $E_{Ri}=10^{18}$ eV, binary mixtures of proton and iron primaries,
and sample size for 5 years of observation. \label{biasML}}
\end{figure}

Hereafter only Sibyll 2.3c will be considered as the reference model, i.e.~HM$_{ref}$ = Sibyll 2.3c. 
Fig.~\ref{biasML} shows $\delta_{c_p}$ (top) and $\Delta \sigma_{c_p}$ (bottom) for different values 
of $\varepsilon$ (see section \ref{SecSim}), i.e.~HM = *Sibyll 2.3c. The case corresponding to 
HM = EPOS-LHC is also shown in the figure. All cases correspond to $E_{Ri}=10^{18}$ eV and $N$ for 
5 years of observation. Note that $\varepsilon = 0$ corresponds to the case in which HM = HM $_{ref}$,
then $\Delta \sigma_{c_p}(\varepsilon = 0)=0$ by definition and $\delta_{c_p}(\varepsilon = 0)=0$ 
since the ML method does not produce bias. Note also that $\delta_{c_p}$ is greater than zero since 
$\varepsilon > 0$. The same applies for the case of EPOS-LHC whose $\langle N_\mu \rangle$ values 
are greater than those of Sibyll 2.3c (Fig.~\ref{MF} top). 

From Fig.~\ref{biasML} it can also be seen that as $\varepsilon$ approaches zero, the curve 
$\delta_{c_p}$ ($\Delta \sigma_{c_p}$) becomes more symmetric (antisymmetric) around $c_p = 0.5$. 
The same can be said for $\varepsilon  < 0$ cases (not shown in the figure) since 
$\delta_{c_p}(\varepsilon) = -\delta_{c_p}(-\varepsilon)$ and 
$\Delta \sigma_{c_p}(\varepsilon) = -\Delta \sigma_{c_p}(-\varepsilon)$ when $|\varepsilon| \ll 1$.
The small differences between the number of muons at ground predicted by post-LHC HEHIMs and therefore, 
between their $\textrm{MF}_{\theta}$, allows us to use any of them as a reference, obtaining similar absolute 
values of $\delta_{c_p}$ and $\Delta \sigma_{c_p}$. From the figure it can also be seen that 
$|\delta_{c_p}|$ and $|\Delta \sigma_{c_p}|$ increase strongly with $\varepsilon$, indicating that 
slight differences between the number of muons predicted by different HEHIMs have a significant impact 
on composition determination (see for instance \cite{Prado:16}).

It is worth mentioning that the values of $\delta_{c_p}$, $\sigma_{\hat{c}_p}$ and $\Delta \sigma_{c_p}$ depend on the statistical
method used to estimate $c_p$. However, impacts of the same order are expected for other methods that make use of the number of muons 
as a mass sensitive parameter. As an example, let us consider the case in which the sample mean of the measured $\widetilde{N}_\mu$
is used to estimate $c_p$. To simplify the calculation let us assume that the energy uncertainty is negligible. Under this assumption 
the following expression for the bias, at order one in $\varepsilon$, is obtained,
\begin{equation}
\delta_{c_p} \cong \varepsilon \left[\frac{ \langle \widetilde{N}_\mu \rangle_{Fe}}{ \langle \widetilde{N}_\mu \rangle_{Fe} - %
\langle \widetilde{N}_\mu \rangle_{p}} - c_{p} \right]
\end{equation}
where $\langle \widetilde{N}_\mu \rangle_{p}$ and $\langle \widetilde{N}_\mu \rangle_{Fe}$ are the mean values of $\widetilde{N}_\mu$
corresponding to proton and Iron primaries for the reference model. Assuming that $\langle \widetilde{N}_\mu \rangle_{p}=27$ and 
$\langle \widetilde{N}_\mu \rangle_{Fe}=42$, the values obtained for Sibyll 2.3c at $\theta=30^\circ$, the median of the 
$\sin(\theta)\cos(\theta)$ distribution for $\theta \in [0^\circ,45^\circ]$ (see Fig.~\ref{MF}), the bias obtained for $\varepsilon=0.09$
decrease from $\sim 0.25$ at $c_p=0$ to $\sim 0.16$ at $c_p=1$, which is even larger than the one obtained for the ML method.

\subsection{Current and future HEHIMs features}
\label{sec:currvsfut}

An analysis performed by the Pierre Auger Observatory indicates that, in the best case, the number of muons measured in the energy range from $10^{18.8}$ to $10^{19.2}$ eV and for $0-60^{\circ}$ inclined showers differs from HEHIMs predictions by a factor $R = 1.33 \pm 0.16$ \cite{Aab:16}. At lower energies, between $10^{17.5}$ and $10^{18}$ eV, the AMIGA data show a muon deficit in simulations of $38\%$ for EPOS-LHC and $50-53\%$ for QGSJETII-04 \cite{Muller:18}. This deficit in the muon content of the showers has been observed also by other experiments in different energy ranges (see Ref.~\cite{Dembinski:19} and references there in). The experimental data show that this deficit increases with primary energy. It is not yet known whether this discrepancy in the number of muons could be indicative of the beginning of some new phenomenon in hadronic interactions at ultra high energies \cite{Farrar:13, Alvarez:13} or can be explained by some incorrectly modeled features of hadronic interactions even at low energy \cite{Dres:04, Maris:09}, as for example, baryon-antibaryon pair production \cite{Pierog:08, Pierog:13} or resonance mesons \cite{Dres:08}. 

The experimental data also show that the muon deficit observed increases with $\sec(\theta)$. This behavior is observed at low energies (between $10^{16.3}$ and $10^{17}$ eV) by the KASCADE-Grande Collaboration \cite{Kascade} and at higher energies (between $10^{18.8}$ and $10^{19.2}$ eV) by Auger \cite{Aab:16}. This behavior can be partially explained by the fact that the experimental value of the attenuation length of the number of muons in the atmosphere is greater than the values obtained from Monte Carlo simulations of the showers \cite{Kascade}. This implies that the observed air showers attenuate more slowly in the atmosphere than the simulated ones. The disagreement on the attenuation length between Monte Carlo predictions and the experimental measurements most likely originates from muon prediction deficiencies of the HEHIMs \cite{Kascade}. The uncertainty in the shape of the muon lateral distribution function employed to reconstruct the EAS data also contributes to this discrepancy, but it is not the principal effect \cite{Kascade}. 

The produced number of muons increases with a small power of the mass number and almost linearly with 
the primary energy. This behavior can be understood in terms of the Heitler-Matthews model of hadronic 
air showers \cite{Matthews:05}, which predicts that the mean value of the total number of muons produced
in a shower is $\langle N_\mu^T \rangle(E,A) = A \left[ E/(A\ \xi_c) \right]^{\beta}$, where $\xi_c$ 
is the critical energy at which charged pions decay into muons. Simulations with post-LHC HEHIMs show 
that $\beta \simeq 0.915-0.928$ \cite{Pierog:17, Prado:16}. Note from Fig.~\ref{MF} (top) that the 
ratio $\langle N_\mu \rangle^{\theta,E}_{Fe}/\langle N_\mu \rangle^{\theta,E}_{p}$, evaluated at 
$E_0 = 10^{18}$ eV, is greater than $\left(56/1\right)^{1-\beta}$. This increment occurs because the 
energy spectrum of muons is harder for iron showers than for proton ones \cite{Espadanal:16}. In the 
same way, since the muon counters are buried underground, the soil attenuates in a greater proportion 
the muons from proton primaries. Although there is still a large uncertainty related to the energy
spectrum of the produced muons \cite{Pierog:17}, in Ref. \cite{Espadanal:16} it was found that by 
changing the energy spectrum by an amount consistent with the difference between current HEHIMs, the
number of muons at ground for the same $\sec(\theta)$ changes by the same factor for all primaries 
and hadronic models. This factor increases (decreases) with $\sec(\theta)$ if the muon energy spectrum 
is hardened (softened). This behavior occurs because at larger zenith angles, muons travel, on average,
larger distances before reaching ground, making the decay of low energy muons more important 
\cite{Espadanal:16}.

On the other hand, the number of muons can change under variation of several important features of 
the hadronic interactions i.e., hadronic particle production cross sections, multiplicity, elasticity 
and, in particular, pion charge-ratio \cite{Ulrich:11}. Fluctuations in the number of muons can also 
change under a modification of these interaction features, being especially sensitive to elasticity 
\cite{Ulrich:11}. The Pierre Auger Observatory also found no evidence of a larger event-to-event 
variance in the ground signal for fixed $X_{max}$ than the one predicted by current HEHIMs \cite{Aab:16}. 
This suggests that the muon deficit cannot be attributed to an exotic phenomenon producing a very large 
muon signal in a fraction of events only \cite{Aab:16}, such as a high rate production of microscopic 
black holes \cite{Feng:02}.

\begin{figure}[t!]
\centering
\setlength{\abovecaptionskip}{0pt}
\includegraphics[width=\linewidth]{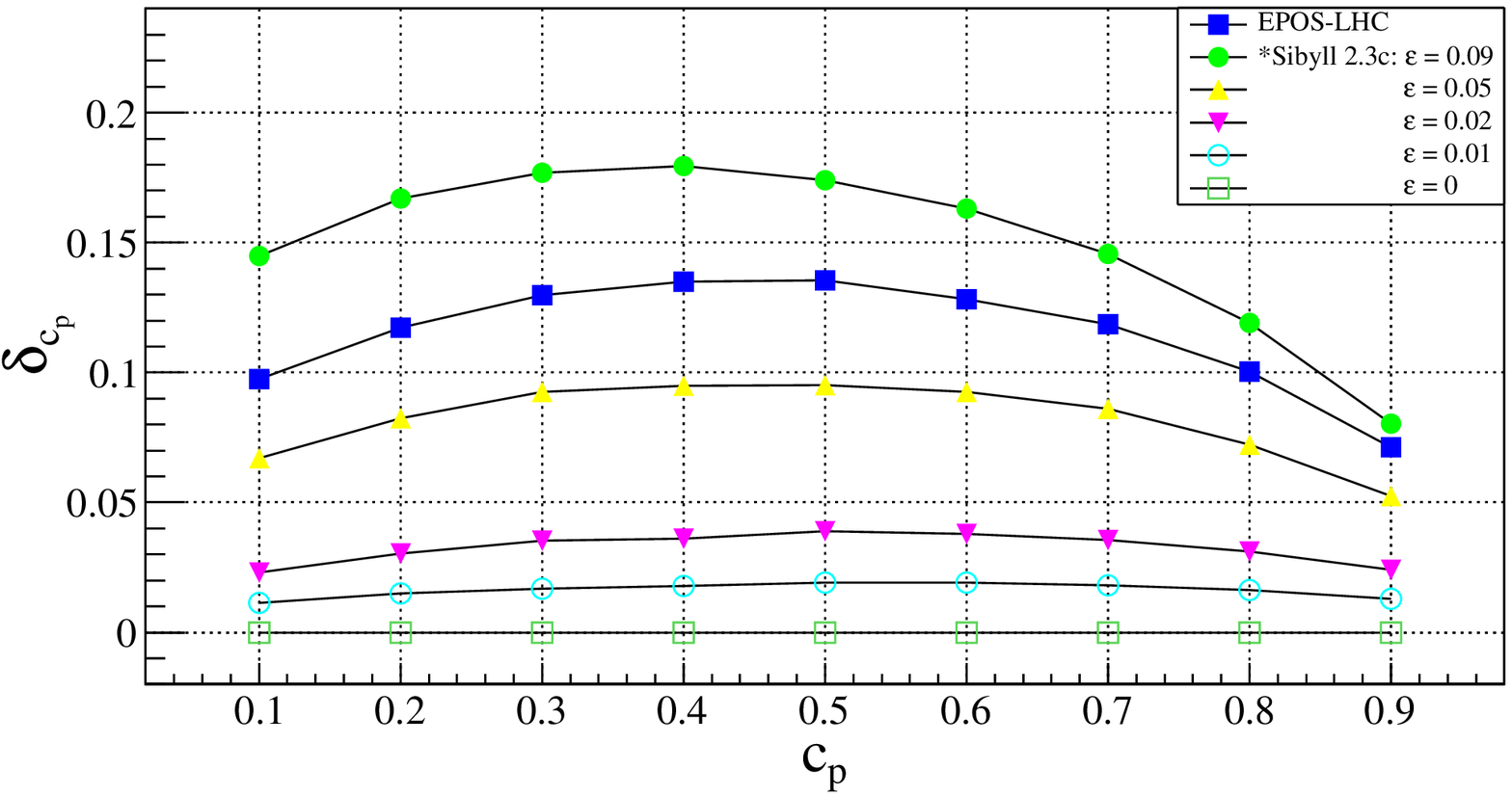}
\includegraphics[width=\linewidth]{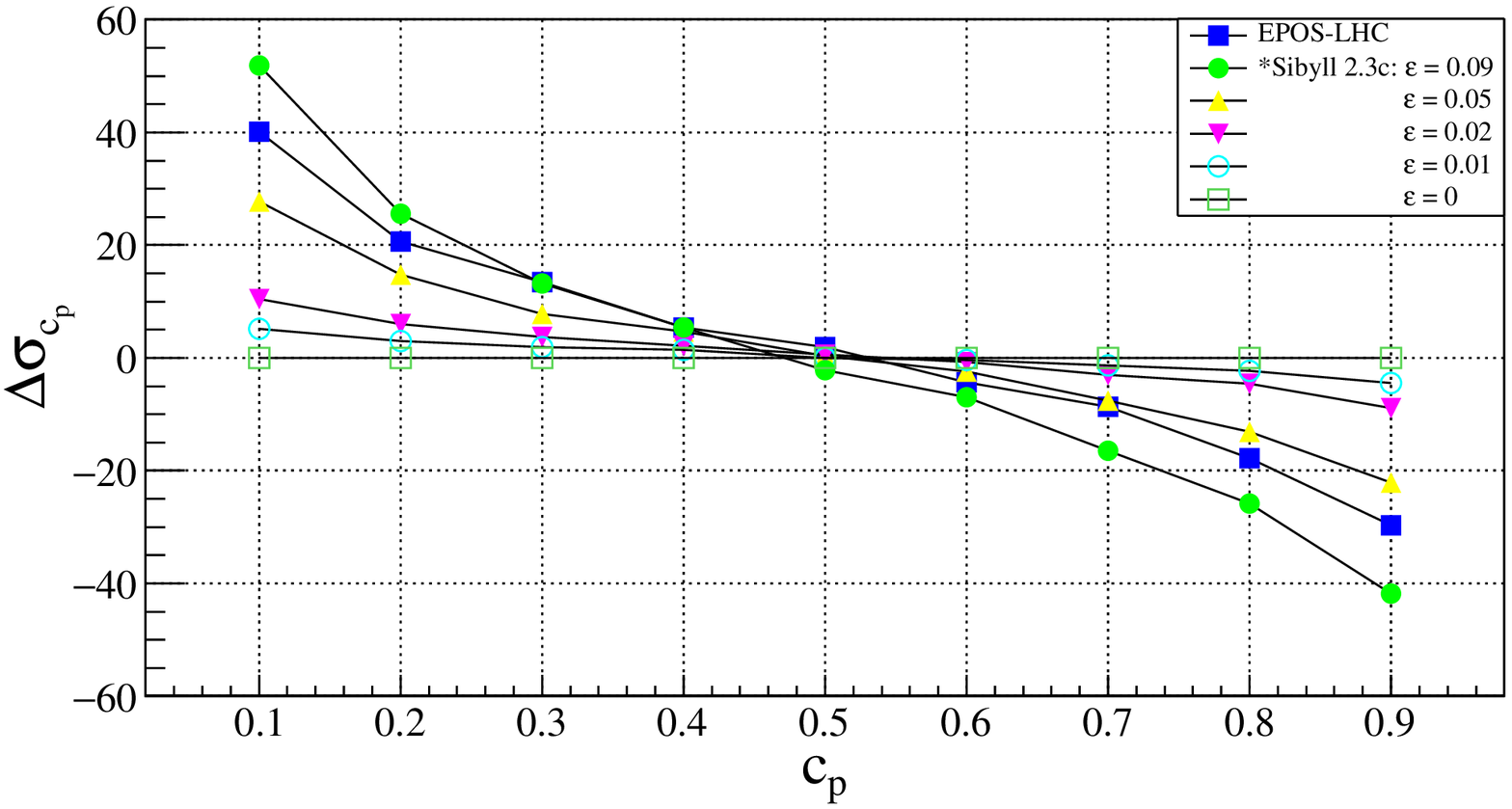}
\caption{$\delta_{c_p}$ (top) and $\Delta \sigma_{c_p}$ (bottom) versus $c_p$ for the ML method and 
for different values of $\varepsilon$ (see section \ref{SecSim}). The case corresponding to HM=EPOS-LHC 
is also shown. All cases correspond to $E_{Ri}=10^{18}$ eV, sample size for 5 years of observation, 
and $R(E,\sec(\theta))=1.4$. 
\label{biasML_14}}
\end{figure}

\begin{figure}[t!]
\centering
\setlength{\abovecaptionskip}{0pt}
\includegraphics[width=\linewidth]{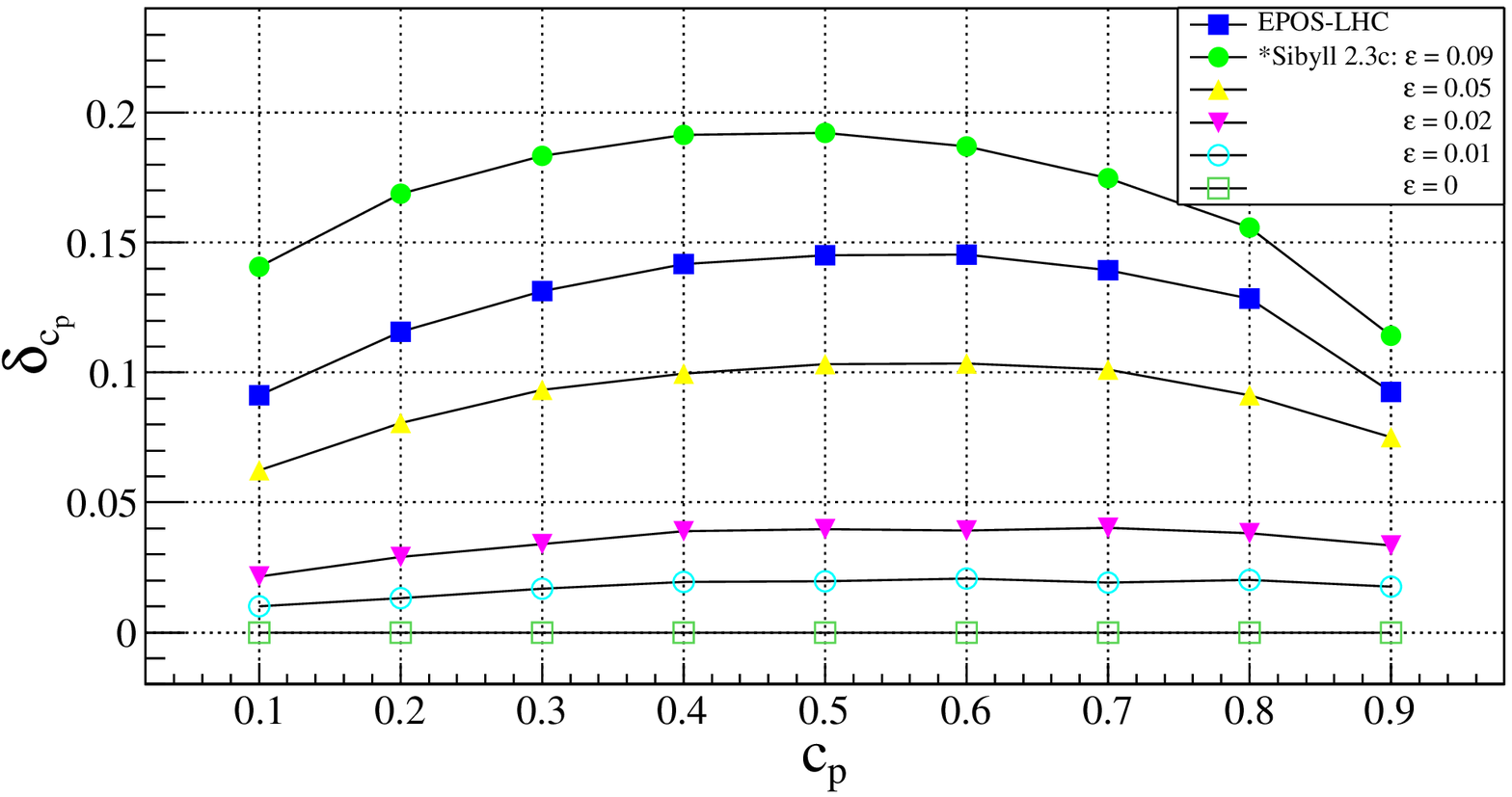}
\includegraphics[width=\linewidth]{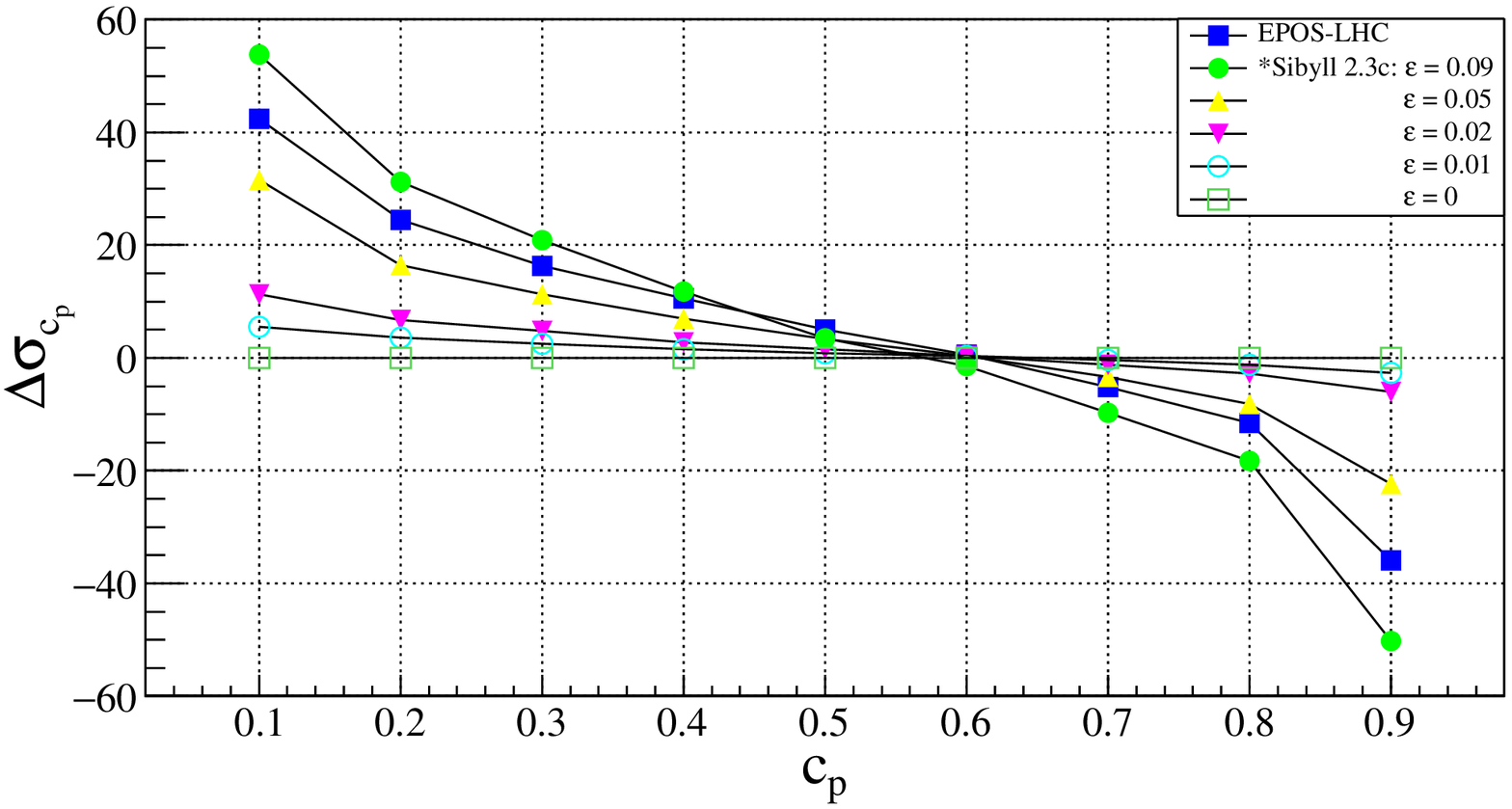}
\caption{$\delta_{c_p}$ (top) and $\Delta \sigma_{c_p}$ (bottom) versus $c_p$ for the ML method and 
for different values of $\varepsilon$ (see section \ref{SecSim}). The case corresponding to HM=EPOS-LHC 
is also shown. All cases correspond to $R(E,\sec(\theta)) = 1.4$, but with the $\sigma_{sh}[N_\mu]^{\theta,E}_{A}$ 
values multiplied by a factor 2 (for both, Sibyll 2.3c and EPOS-LHC). 
\label{biasML_14y2}}
\end{figure}

It is expected that the observed muon deficit can be reduced in the next generation of HEHIMs, at 
least for $0-60^{\circ}$ inclined showers \cite{Pierog:17}. Therefore, assuming that the current mean 
values $\langle N_\mu \rangle^{\theta,E}_{A}$ will be increased by some factor, $R(E,\sec(\theta))$ 
(almost independent of $A$), the difference 
$\langle N_\mu \rangle^{\theta,E}_{A_2} - \langle N_\mu \rangle^{\theta,E}_{A_1}$ will be increased by 
the same factor. Although it is not possible to know whether $\sigma_{sh}[N_\mu]^{\theta,E}_{A}$ will 
increase, decrease or remain unchanged, $\sigma[\widetilde{N}_\mu]^{\theta,E}_A$ will increase because 
it is dominated by $\langle N_\mu \rangle^{\theta,E}_{A}$ (see Eq.~(\ref{sig_A})). Then, the 
$\textrm{MF}_{\theta}$ will suffer only a slight modification, having a similar impact on composition 
determination under the same relative differences between HEHIMs, or eventually, between a given 
HEHIM and the experimental data. Fig.~\ref{biasML_14} shows $\delta_{c_p}$ (top) and $\Delta \sigma_{c_p}$ 
(bottom) for the same cases of Fig.~\ref{biasML} but with the $\langle N_\mu \rangle^{\theta,E}_{A}$ 
values multiplied by a hypothetical (naive) factor $R(E,\sec(\theta)) = 1.4$. In the same way 
Fig.~\ref{biasML_14y2} shows $\delta_{c_p}$ (top) and $\Delta \sigma_{c_p}$ (bottom) for the same cases 
of Fig.~\ref{biasML_14} with $R(E,\sec(\theta)) = 1.4$ but with the $\sigma_{sh}[N_\mu]^{\theta,E}_{A}$ 
values multiplied by a factor 2. It can be seen that $|\delta_{c_p}|$ and $|\Delta \sigma_{c_p}|$ of 
Figs.~\ref{biasML_14} and \ref{biasML_14y2} are similar to the ones corresponding to Fig.~\ref{biasML}.  
Therefore, these results suggest that future HEHIMs will have a similar impact on composition determination 
than current ones.

\subsection{Application to a simplified case}
\label{sec:simpcase}

In this section the impact of the HEHIMs in composition analyses is studied in the energy range from
$10^{17.5}$ and $10^{18.5}$ eV, assuming a binary mixture of protons and nitrogen nuclei. This assumption
is based on the results obtained in Ref.~\cite{Bellido}. In that work a sample of 42466 events recorded 
by the Pierre Auger Observatory is used to obtain the $X_{max}$ distributions in energy bins of 
$\Delta \log(E/\textrm{eV})=0.1$, ranging from $10^{17.2}$ eV to $10^{19.6}$ eV. The experimental 
distributions are fitted considering the $X_{max}$ distributions obtained for Sibyll 2.3c, EPOS-LHC, 
and QGSJETII-04, including the detector effects. The composition fractions are estimated assuming four 
elemental primary groups: proton, helium, nitrogen, and iron. In the energy range from $10^{17.5}$ to 
$10^{18.5}$ eV protons and nitrogen nuclei turn out to be the most abundant nuclear species when 
Sibyll 2.3c is considered to analyze the $X_{max}$ Auger data. Note that the fraction of helium is 
also appreciable for energies greater than $10^{18}$ eV but with large error bars. 

\begin{figure}[t]
\centering
\setlength{\abovecaptionskip}{0pt}
\includegraphics[width=\linewidth]{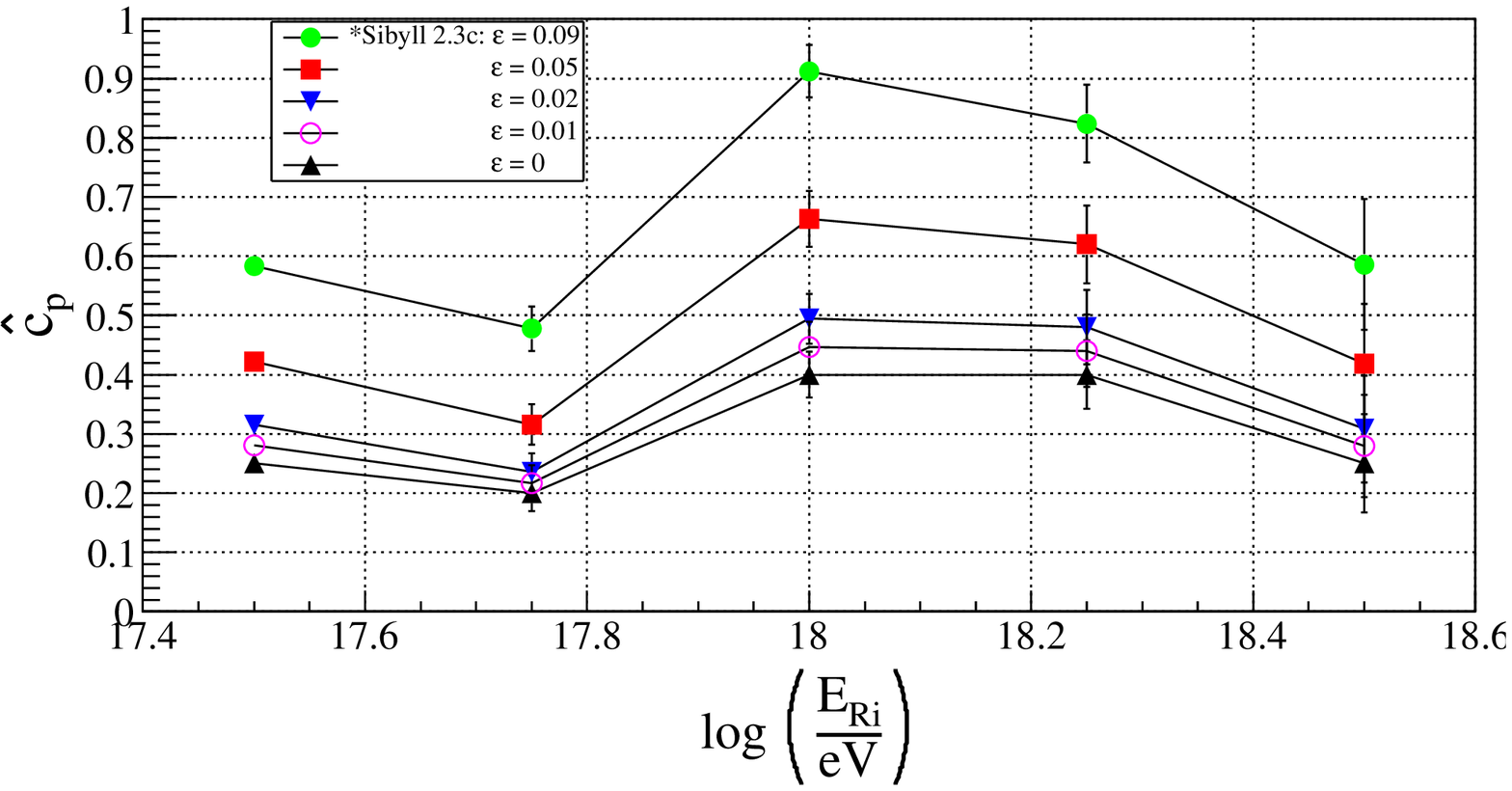}
\includegraphics[width=\linewidth]{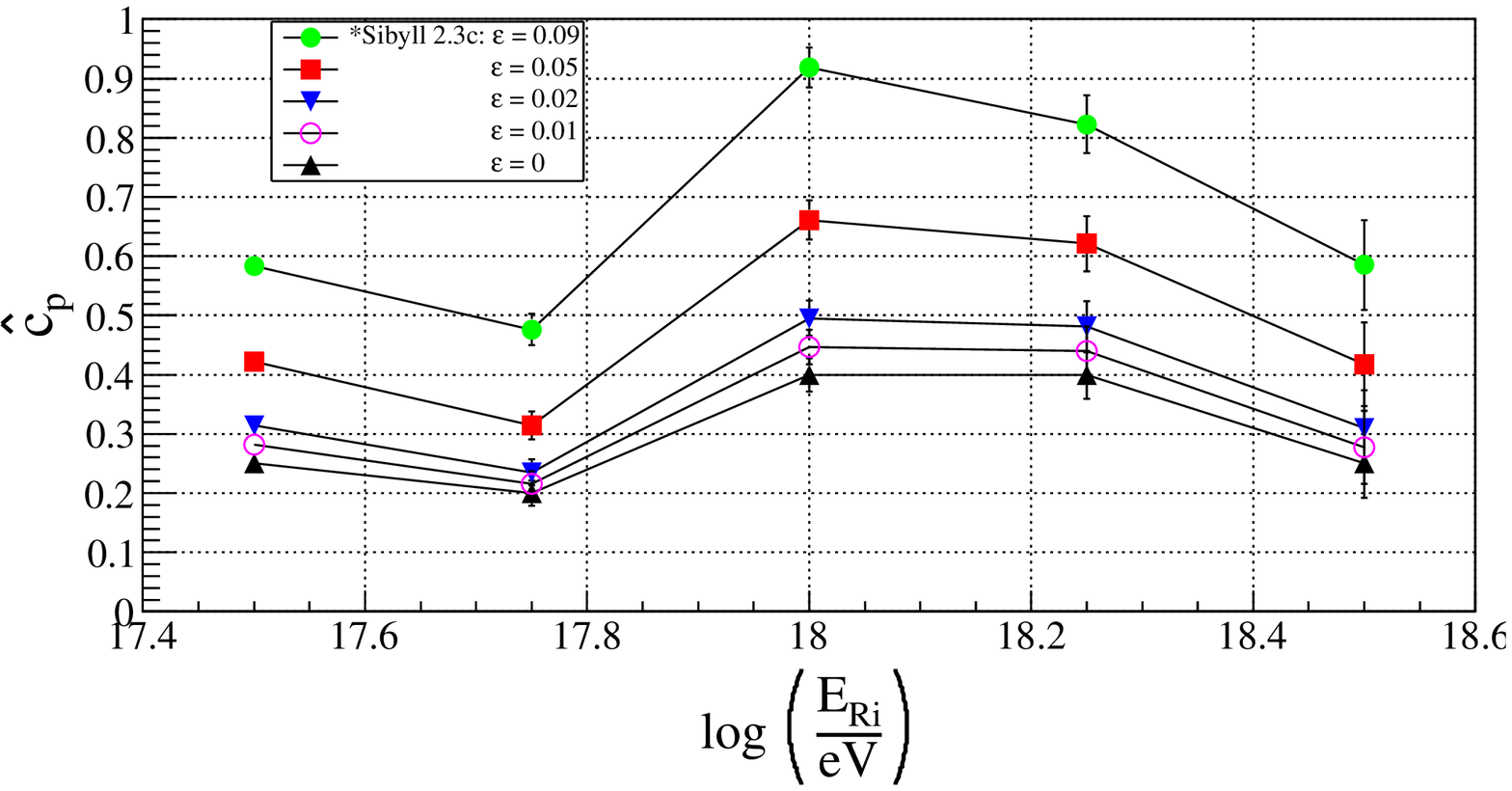}
\caption{$\hat{c}_p$ as a function of $E_{Ri}$ for different values of $\varepsilon$ (see section
\ref{SecSim}) and sample size corresponding to 5 (top) and 10 (bottom) years of observation. The 
error bars correspond to $\sigma_{\hat{c}_p}$. $\sigma[\epsilon]$ for nitrogen is approximated by 
the average between the ones corresponding to proton and iron primaries. 
\label{Cph01_E}}
\end{figure}

The top panel of Fig.~\ref{Cph01_E} shows the inferred $\hat{c}_p$ values with their statistical 
uncertainties for sample size $N$ corresponding to 5 years of observation and Sibyll 2.3c as the 
reference HEHIM, obtained for different values of $\varepsilon$ in the energy range under consideration.
The relative error of the reconstructed number of muons, $\sigma[\epsilon]$, used for nitrogen is 
approximated by the average between the ones corresponding to proton and iron primaries. The values 
of $c_{p}(E_{Ri})$ considered as input are extracted from the Fig.~6 of Ref.~\cite{Bellido}. The error 
bars correspond to $\sigma_{\hat{c}_p}$. Note that 
$\delta_{c_p}(E_{Ri} = 10^{18}\ \textrm{eV}, \varepsilon)$ are larger than those of Fig.~\ref{biasML} 
(top), this is due to the fact that the merit factor for proton and nitrogen is smaller than the one 
corresponding to proton and iron.
The bottom panel of Fig.~\ref{Cph01_E} shows the $\hat{c}_p$ values for the same case shown at the top 
one but for sample size $N$ corresponding to 10 years of observation. As expected, the error bars of 
the reconstructed $c_{p}$ are smaller than those at the top panel.

\begin{figure}[t]
\centering
\setlength{\abovecaptionskip}{0pt}
\includegraphics[width=\linewidth]{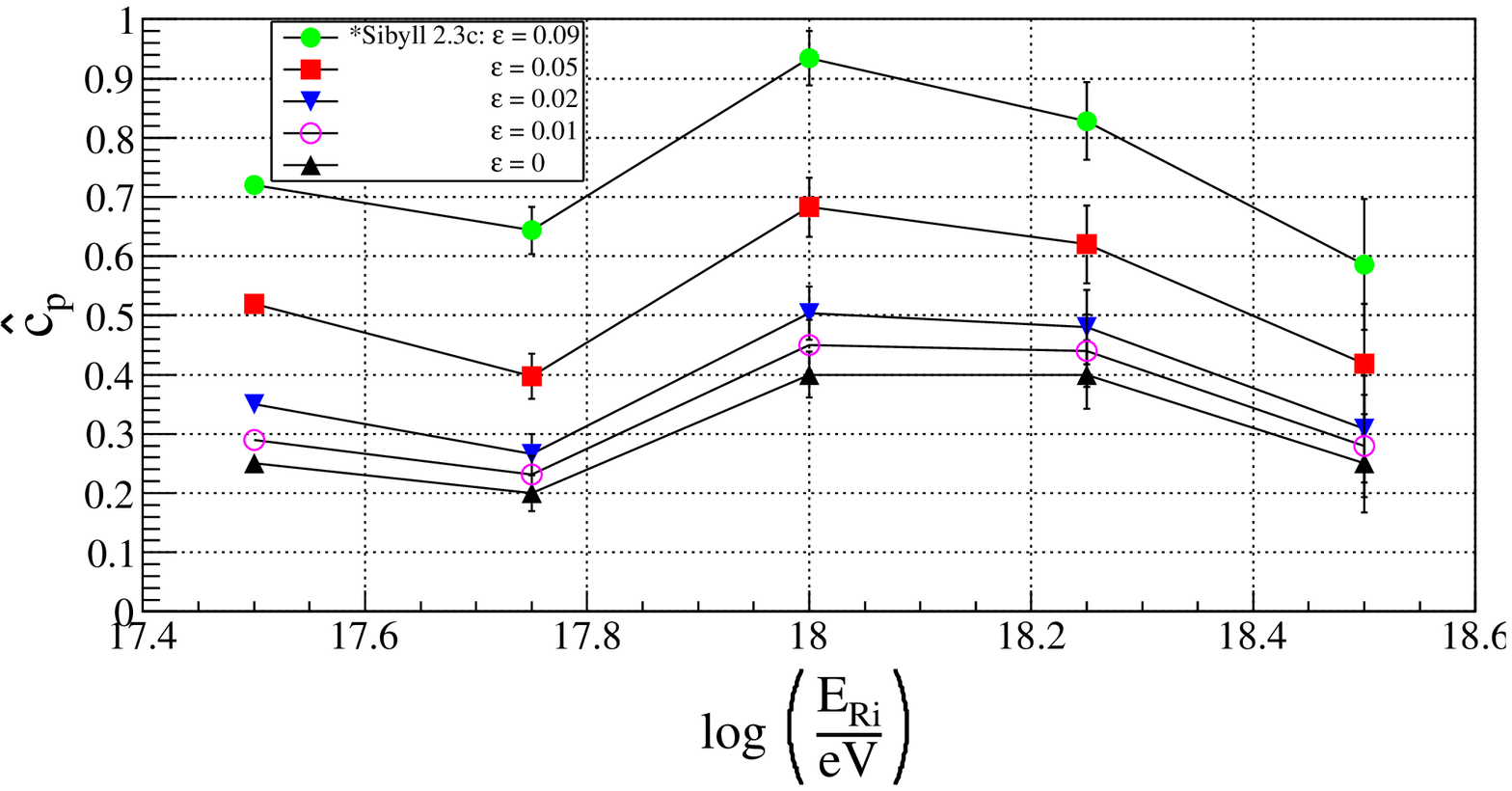}
\includegraphics[width=\linewidth]{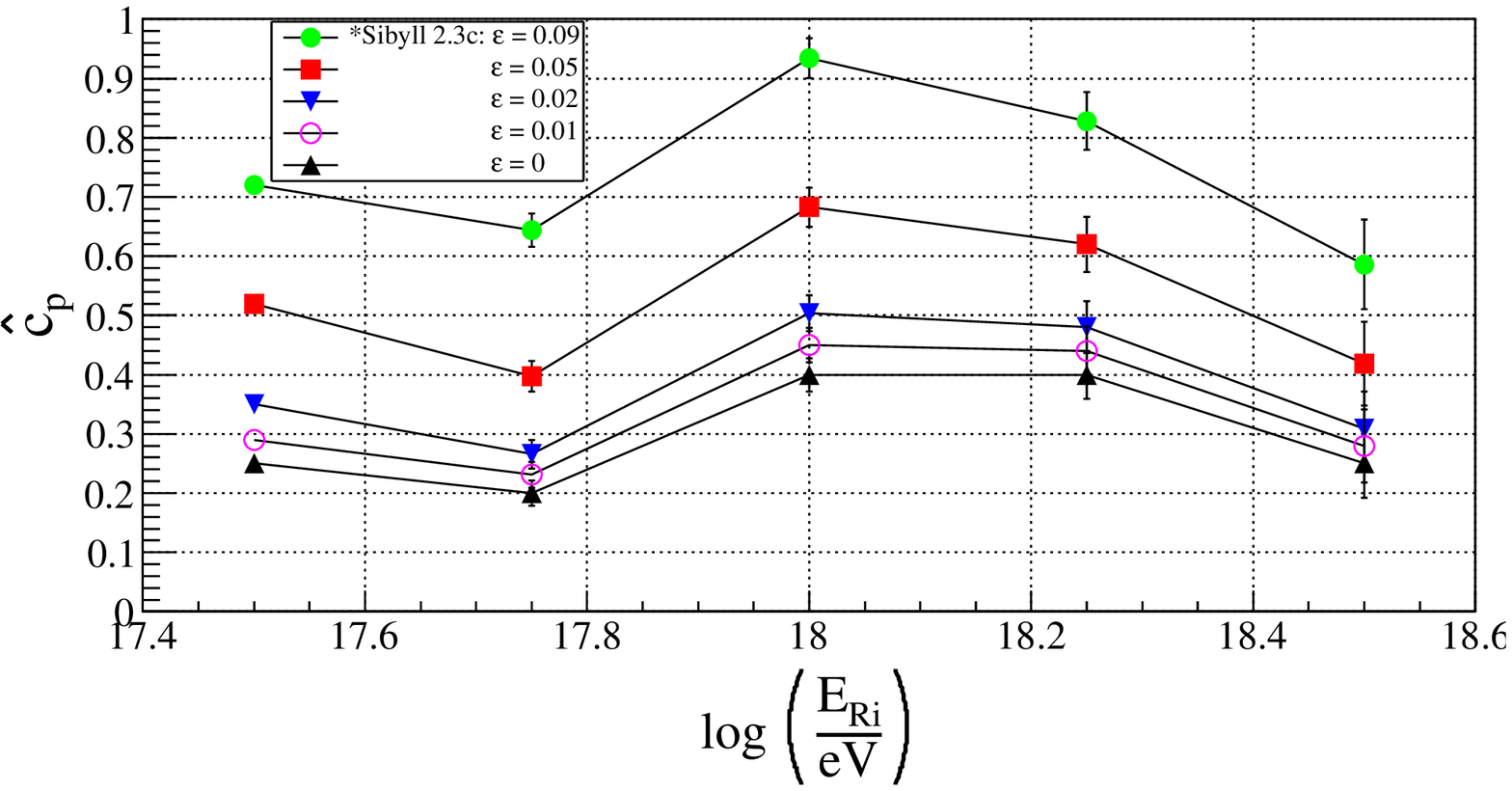}
\caption{$\hat{c}_p$ as a function of $E_{Ri}$ for different values of $\varepsilon$ (see section
\ref{SecSim}) and sample size corresponding to 5 (top) and 10 (bottom) years of observation. The 
error bars correspond to $\sigma_{\hat{c}_p}$. $\sigma[\epsilon]$ used for nitrogen is the same as 
the one obtained for proton primaries (conservative case). 
\label{Cph02_E}}
\end{figure}

Figure \ref{Cph02_E} shows the inferred $\hat{c}_p$ values for the same cases as in Fig.~\ref{Cph01_E} 
but using the same $\sigma[\epsilon]$ for nitrogen as for proton (conservative case). As expected, the 
biases and the error bars in Fig.~\ref{Cph02_E} are larger than those of Fig.~\ref{Cph01_E} (especially 
for $\varepsilon \gtrsim 0.05$). However, they are very similar for energies $\geq 10^{18}$ eV. This can 
be explained from Fig.~\ref{rec} where it is seen that for energies above $10^{18}$ eV $\sigma[\epsilon]$ 
of protons is larger than $\sigma[\epsilon]$ of iron nuclei in less than 3\%.
From Figs.~\ref{Cph01_E} and \ref{Cph02_E} it can also be seen that to ensure $|\delta_{c_p}| \lesssim 0.05$, 
the differences between $\langle N_\mu \rangle^{\theta,E}_{A}$ of different HEHIMs must be smaller than $2\%$. 
It is well known that the mass composition determination obtained from the $X_{max}$ parameter 
also depends on the HEHIMs used in the analyses \cite{Bellido, Abreu:13, Aab:14}. Comparing the 
results obtained in this work with the ones obtained in Ref.~\cite{Bellido} at $E_{Ri} = 10^{18}$ eV 
(energy at which the mass composition obtained by using $X_{max}$ corresponds basically to a binary
mixture of protons and nitrogen nuclei), it can be seen that 
$\delta_{c_p}(E_{Ri} = 10^{18}\ \textrm{eV}, \varepsilon=0.05)$ is of the order of the bias found in 
the analysis based on the $X_{max}$ parameter when Sibyll 2.3c and EPOS-LHC are used to estimate the 
proton abundance. Therefore, it indicates that the systematic uncertainties introduced by the use
of different HEHIMs on composition analyses based on the $N_\mu$ parameter are of the same order 
as the ones based on the $X_{max}$ parameter. 

In summary it is found that small differences between the predicted values of 
$\langle N_\mu \rangle^{\theta,E}_{A}$ and $\sigma_{sh}[N_\mu]^{\theta,E}_{A}$ obtained when different 
HEHIMs are considered have a significant impact on composition determination. It is worth mentioning 
that in recent years different mass composition methods that seem to have a reduced dependence on the 
assumed HEHIM have been developed. In Ref.~\cite{Blaess:18} a method of this type, which is based on 
the $X_{max}$ distributions, is presented. This method is based on parametrizations of the $X_{max}$ 
distributions obtained from simulations in which the normalization levels of the mean value and the 
standard deviation of $X_{max}$ are determined from experimental data. In this way the influence of 
HEHIMs on composition analyses is reduced. In Ref.~\cite{Younk:12} a new method, based on the correlation 
between $X_{max}$ and the number of muons in air showers, is introduced. The purpose of this method is to 
determine whether the mass composition is pure or mixed. A similar method is used by Auger to study the 
composition in the ankle region \cite{Aab:16-2}. In this case the correlation between $X_{max}$ and 
$S$(1000), the signal of the Cherenkov detectors at 1000 m from the shower axis, is considered. The 
use of $S$(1000) is based on the fact that for $\theta=20-60^{\circ}$ the muon component represent the 
$40-90$\% of the total signal at 1000 m from the shower axis \cite{Kegl:13}. The results obtained in 
Ref.~\cite{Aab:16-2} are robust with respect to experimental systematic uncertainties and to the 
details of the hadronic interactions. Therefore, it is expected that similar methods based on the 
combination of the number of muons with other mass sensitive parameters can be developed in order to 
reduce the dependence of the composition analyses on HEHIMs.

\section{Conclusions}
\label{sec:conclusions}

We have presented a maximum likelihood method to perform mass composition analyses based on the number 
of muons  at a given distance to the shower axis measured by underground muon detectors. This method
includes the dependency of the muon number with the zenith angle of the showers. All the studies
have been done from simulations, which include the effects introduced by the detectors and the 
reconstruction methods. The shape of the energy spectrum in combination with the uncertainties in the 
reconstruction of the primary energy has also been taken into account. The proton abundance has been 
estimated with a good statistical resolution, reaffirming that the muon content of the shower is one 
of the best indicators of the primary mass.

We have also studied in detail the impact of the use of different high energy hadronic interaction 
models in the composition analyses performed by using the method developed. The biases introduced 
by the differences on the prediction of that models resulted to be the dominant uncertainties on
the composition determination. The development of composition methods with a reduced influence of
the high energy hadronic interaction models are required in order to reduce these important
systematic uncertainties.

\begin{acknowledgments}
A. D. S. and A. E. are member of the Carrera del Investigador Cient\'ifico 
of CONICET, Argentina. This work is supported by ANPCyT PICT-2015-2752, 
Argentina. The authors thank the members of the Pierre Auger Collaboration 
for useful discussions.
\end{acknowledgments}

\end{document}